\documentclass[aps, prd, preprint, nofootinbib, amsfonts, floatfix]{revtex4}

\usepackage{amsmath}  
\usepackage{amsfonts} 
\usepackage{graphicx} 
\usepackage[title]{appendix}
\usepackage{hyperref}
\usepackage[capitalise]{cleveref}
\usepackage{physics}
\usepackage{qcircuit}
\usepackage{tikz}
\usepackage{fancyvrb}
\usetikzlibrary{graphs,shapes.misc} 
\tikzset{terminal/.style={
                      rectangle,minimum size=6mm,rounded corners=3mm,
                      very thick,draw=orange!70!black!50,
                      top color=white,bottom color=orange!20,
                      font=\ttfamily},
         nonterminal/.style={
         rectangle,
          minimum size=6mm,
          very thick,
          draw=red!30!black!70,         
          top color=white,              
          bottom color=red!70!black!30, 
          font=\itshape
    }}

\definecolor{lesbian-red}{HTML}{D62800}
\definecolor{lesbian-dark-orange}{HTML}{EF7627}
\definecolor{lesbian-orange}{HTML}{FF9B56}
\definecolor{lesbian-pink}{HTML}{D462A6}
\definecolor{lesbian-purple}{HTML}{A40062}

\begin{document}


\title{Building intuition for Quantum Algorithms using the Aharonov-Bohm effect}

\author{Anna Liv Paludan Bjerregaard}
\author{Kim Splittorff} 
\email{split@nbi.ku.dk  (corresponding author)}
\affiliation{${\ }$ NNF Quantum Computing Programme, \\ Niels Bohr Institute, University of Copenhagen, Blegdamsvej 17, 2100 Østerbro, Copenhagen, Denmark} 



\date{\today}

\begin{abstract}
    Quantum computers operate in a manner so different from classical computers that our intuitions for understanding the process of quantum computation is challenged. This not only makes it hard to develop new quantum algorithms, it also makes quantum computation a difficult subject for students to enter. Here we show that the time evolution of a particle on a ring with a coil through performs the key quantum algorithm known as  Quantum Phase Estimation.     
    The example uses the Aharonov-Bohm effect, and the aim is to help students build a physical intuition for quantum algorithms.
\end{abstract}

\maketitle 

\section{Introduction}

A quantum computer is a quantum system and a quantum computation is a physical process in this quantum system \cite{qubitguide}. The quantum computer can thereby in some cases solve a numerical problem more efficiently than a classical computer \cite{NielsenChuang,qubitguide,Aaronson,Preskill}. The potential for applications spans many areas of physics, chemistry and life science \cite{QPEforChem}. However, Peter Shor has noted \cite{ShorWhyNoAlg}  ``that quantum computers operate in a manner so different from classical computers that our techniques for designing algorithms and our intuitions for understanding the process of computation no longer work.'' This lack of intuition not only makes it challenging to develop new quantum algorithms, it also makes quantum computation a difficult subject for students to enter.

Algorithms for quantum computers are often introduced in purely mathematical terms, describing how to manipulate quantum states to obtain a desired numerical result extremely efficiently. Such descriptions are very useful, but perhaps lack appeal to the kind of intuition Peter Shor refers to. 
 The aim of this article is to help students build an intuition for quantum algorithms, by providing an example of a quantum system which solves a numerical problem more efficiently than a classical computer. 
Starting from the textbook example of the Aharonov-Bohm effect for a particle on a ring \cite{Griffiths,Sakurai}, we will show that the time evolution of the state of the particle exactly corresponds to the Quantum Phase Estimation (QPE) \cite{kitaevQPE,QuantAlgsRevisit,LloydQPE} algorithm. In other words, we will show that the particle on the ring is a quantum computer capable of performing the QPE algorithm. 
QPE is an essential subroutine in many other quantum algorithms, such as Shor's algorithm for factoring primes\cite{Shor} and algorithms for estimating ground state energies in quantum systems \cite{QPEforChem,kitaevQPE}. As such, gaining a physical intuition for QPE opens the door for students to study a wide array of quantum algorithms.

Several important quantum algorithms have been shown to have physical analogues \cite{SplitQPEAB,GroverClassical,GroverITE}. 
The link between the Aharonov-Bohm effect for a particle on a ring and the QPE algorithm was established in \cite{SplitQPEAB}. There the physical system was used as a way to study what happens to a quantum computer in the classical limit $\hbar\to0$. In this article we introduce the link established in \cite{SplitQPEAB} in a simplified manner by focusing purely on the topological properties of the system, i.e.~how many times the state of the particle winds around the ring in a given period of time. Moreover, to fully realise the QPE algorithm the non-Abelian Aharonov-Bohm effect is needed and, as an example, we consider how the evolution of the combined position and colour state of a quark on the ring solves a $2\times2$ eigenvalue problem. In addition, this paper aims to give students insight into why quantum computers in some cases can provide immense computational speedup compared to classical methods \cite{Shor,QuantAlgsRevisit,HHL}.
To do so we consider how hard it is classically to simulate the quantum system which makes up the quantum computer: if it was easy to classically simulate the quantum system which makes up the quantum computer, it would not be possible for the quantum computer to have an advantage over classical computers.  To this end we show that classical simulation of the 
particle on a ring scales poorly contrasted with the quantum algorithm it can perform. 

Our hope is that this paper can help students build an intuition for quantum computing, perhaps it can also serve as a guide for how to introduce quantum algorithms to students from a physicists perspective. The paper is organised as follows: \Cref{sec:TheProblem} discusses the eigenvalue problem we wish to solve and how to think about encoding the necessary information into a quantum system, as well as how this quantum system would need to be manipulated in order to read out the correct result. In \cref{sec:ABeffect} we introduce the Aharonov-Bohm effect as a way of achieving this manipulation by time evolution of the quantum system. The example with the quark and the $2\times2$ matrix is given in Section \ref{sec:ABExample}. Then in \cref{sec:QPE} we show how our quantum system implements the exact steps of the QPE algorithm. Lastly in \cref{sec:simul} we exemplify how quantum computers can provide computational advantage for certain problems. To do so we simulate classically the Aharonov-Bohm effect for a particle on a ring, a problem which is seemingly simple, but where the amount of operations scales poorly with the desired precision.

\section{The Problem}\label{sec:TheProblem}

An important part of many numerical calculations is to determine eigenvalues of a given operator. Here we will consider a  unitary operator $U$ (that is an operator for which $U^\dagger U=UU^\dagger=1$) with an eigenvector $\ket{u}$ such that
\begin{align}
    U\ket{u}=e^{i\phi_u}\ket{u}\:.\label{eq:TheProblem}
\end{align}
Note that the eigenvalues of unitary operators are complex phases $e^{i\phi_u}$ (to see this, denote a given eigenvalue of $U$ by $\lambda$, and use that $1=\langle u | u\rangle = \bra{u}U^\dagger U\ket{u}=|\lambda|^2$, such that the eigenvalue $\lambda=e^{i\phi_u}$).

As an example let us consider the unitary $2\times2$ matrix,  $U = \exp(i E_0 \sigma_1/E_R)$, where $\sigma_1$ is the first Pauli matrix, the energy $E_0>0$ and $E_R$ is a reference energy. We are given the eigenvector $\ket{u}=\ket{-}$ of $U$, where $\sigma_1\ket{-}=(-1)\ket{-}$ and our task is to find the phase, $\phi_u$ of the related eigenvalue $e^{i\phi_u}$. (For those not familiar with the matrix structure of the Pauli matrices we have written it out explicitly in Section \ref{sec:ABExample} where we will return to this example.) As we are here dealing with $2\times2$ matrices and since $U$ is on an exponential form, it can be seen that $\phi_u=-E_0/E_R$. It is of course not challenging for classical computers to find eigenvalues of $2\times2$ matrices. Suppose we instead consider a $U$ which is the time evolution operator of a quantum system, then determining an eigenvalue of $U$ corresponds to determining an eigenvalue of the Hamiltonian. If the Hilbert space is large and the Hamiltonian describes complex interactions, determining for example the ground state energy can be extremely challenging on a classical computer.

\subsection{The particle on a ring}
\begin{figure}
    \centering
    \includegraphics[width=0.4\linewidth, trim = 100 100 0 0, clip]{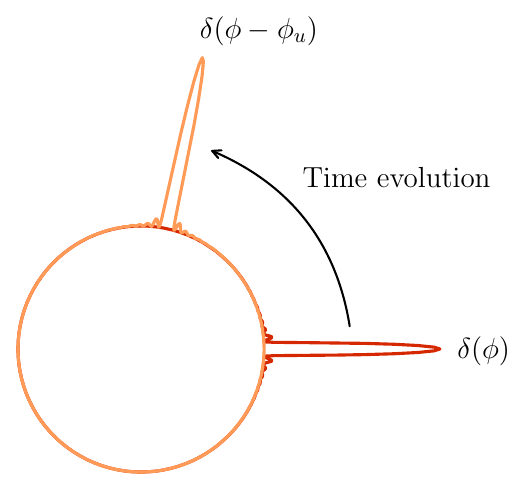}
    \caption{Illustration of moving information in a quantum system by time evolving the position-space probability density for a particle on a ring. Note that ``time evolution'' need not refer to a simple rotation; such a rotation would require knowledge of $\phi_u$.}
    \label{fig:IntroFig}
\end{figure}

Given the quantum state $\ket{u}$ and the operator $U$, our aim is to identify a quantum system (our quantum computer) which allows us to determine the phase of the eigenvalue $\phi_u$. Since $\phi_u$ is an angle, we will consider a particle on a ring of radius $r$
and identify $\phi$ as the angular position of the particle. We wish to find a time evolution which depends on $U$ and takes the particle from a wave function initially localised at $\phi=0$, to one localised at $\phi=\phi_u$,
\begin{align}
    \delta(\phi)\rightarrow\delta(\phi-\phi_u)\:,\label{eq:deltaShift}
\end{align}
as illustrated in \cref{fig:IntroFig}. A measurement of the angular position will then give the desired phase of the eigenvalue $\phi_u$.

Essentially, we have reduced the eigenvalue problem to the problem of moving information around a quantum system. This is the general idea of quantum algorithms: encoding information into quantum systems and manipulating it using the rules of quantum mechanics \cite{NielsenChuang,qubitguide}. However, as stated initially, we do not \emph{know} $\phi_u$ a priori. We need a way to obtain \cref{eq:deltaShift} using only $U$ and $\ket{u}$.

\subsection{Fourier Multipliers}

Our aim is to shift the position-space wave function of a particle on a ring of radius $r$, $\psi(\phi)\rightarrow\psi(\phi-\phi_u)$. Mathematically this can be achieved by multiplying each of its Fourier basis functions $\psi_m(\phi)=\frac{1}{\sqrt{2\pi}}e^{im\phi},  m\in\mathbb{Z}$ by $e^{-im\phi_u}$,
\begin{eqnarray}
    \psi(\phi) & = & \sum_{m=-\infty}^\infty c_m\psi_m(\phi)\\
    \psi(\phi-\phi_u) & = & \sum_{m=-\infty}^\infty c_me^{-im\phi_u} \psi_m(\phi) \label{eq:inverseFourier} \:. 
\end{eqnarray}
The phase $e^{-im\phi_u}$ is called a {\sl Fourier multiplier}. Physically our task is to set up the particle on the ring  (our quantum computer) such that the time evolution realises the Fourier multipliers. If successful the particle will localise at $\phi_u$, such that we can read out $\phi_u$ by measuring the position of the particle. 
Note that we must set up this quantum computation  
without knowing $\phi_u$.


\section{The Aharonov-Bohm Effect}\label{sec:ABeffect}

As we shall now see, the Aharonov-Bohm effect \cite{AharonovBohm} naturally provides the Fourier multiplier, which allows us to shift the particles wave function and obtain $\phi_u$ by measuring $\phi$.

\subsection{The standard Aharonov-Bohm Effect}

The Aharonov-Bohm effect describes how magnetic vector potentials affect the phase of electrically charged particles. More specifically, a particle with electromagnetic charge $q$ traveling along a path $\gamma$ will pick up an Aharonov-Bohm phase given by \cite{PhaseFactors}
\begin{equation}
    \exp(\frac{iq}{\hbar}\int_\gamma\vb{A}\cdot\dd\vb{r})\:.\label{eq:ABphase}
\end{equation}
The standard example of this is a particle moving past a solenoid of infinite length \cite{Griffiths,Sakurai}. Inside the solenoid there is a constant magnetic field in the $\emph{z}$-direction, $\vb{B}=B_0\hat{z}$, but outside the solenoid $\vb{B}=0$, see \cref{fig:ABFourierMult}. The vector potential outside the solenoid in cylindrical coordinates, $(r,\phi,z)$, is $\vb{A}(r) = \frac{\Phi}{2\pi r}\hat{\phi}$, where $\Phi$ is the flux of the $\vb{B}$-field through a cross-section of the solenoid. As suggested by Aharonov and Bohm \cite{AharonovBohm} by performing a sort of ``double slit'' experiment, allowing an electron to pass either way around the solenoid in the $x$-$y$ plane and letting it interfere with itself, one can observe the phase difference $\Delta \phi$, between the two paths, as illustrated in \cref{fig:ABFourierMult}a. 
Note that the Aharonov-Bohm phase is purely geometric effect \cite{AharonovBohm}: as long as the two paths pass on each side of the solenoid the Aharonov-Bohm phase will take the same value given by the total flux of the $\vb{B}$-field through the solenoid.

\subsection{Fourier multipliers from the Aharonov-Bohm phase}

As we now show, when we restrict the particle to move on a ring around the solenoid the Aharonov-Bohm phase gives exactly the Fourier multipliers we need. 

\begin{figure}
    \centering
    \includegraphics[width=0.7\linewidth]{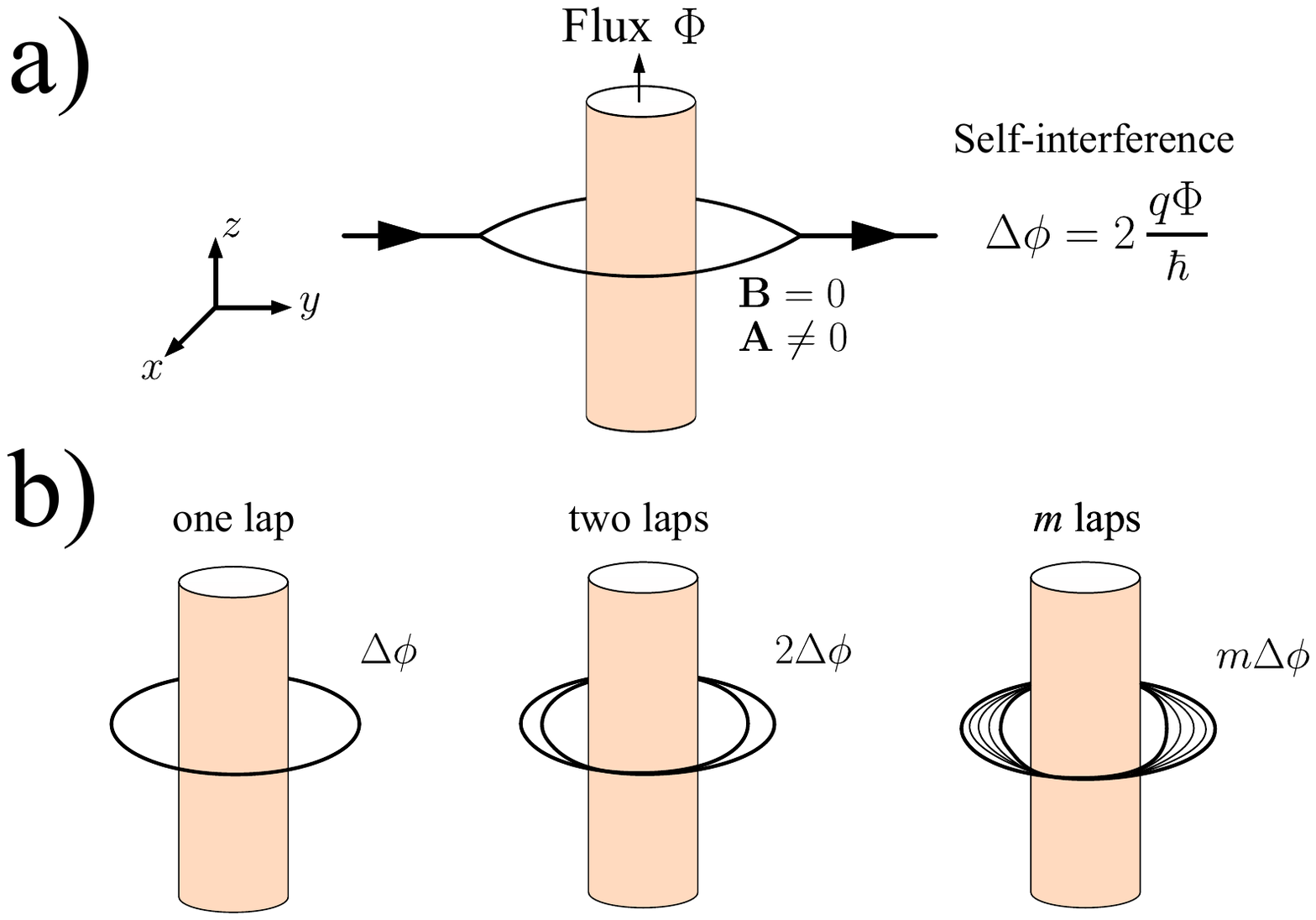}
    \caption{{\bf a)} Example of the Aharonov-Bohm effect. A particle is allowed to pass either way around a solenoid and self-interferes at the other end, each path having picked up a relative phase. {\bf b)} The Aharonov-Bohm effect for constructing Fourier multipliers. Each mode $\psi_m(\phi)$ of the particle travels $m$ laps around the ring in time $t_R$, thereby picking up a phase $m\Delta\phi$.}
    \label{fig:ABFourierMult}
\end{figure}

 We initially localise the particle on then ring, cf.~\cref{eq:deltaShift}, and express the $\delta$-function as an infinite sum over $\psi_m(\phi)$
 \footnote{Note that to obtain the $\delta$-function we sum over an even interval of angular momenta, $\hbar m$, running from $-\hbar l$ to $\hbar l$ and then take $l$ to infinity. The number $l$ is not to be confused with the $l$ quantum number of a 3-dimensional spherical problem.}
\begin{equation}
    \delta(\phi)=\lim_{l\rightarrow\infty}\frac{1}{\sqrt{2l+1}}\sum_{m=-l}^l\psi_m(\phi)\:.\label{eq:initwave}
\end{equation}
Each of these states $\psi_m$ has a well defined angular momentum, $\hbar m$, and an associated angular velocity, $\omega = \hbar m/m_qr^2$. Here $m_q$ is the mass of the particle and $r$ is the radius of the ring. Thus if we wait the time it takes for the $m=1$ mode to complete a full lap around the ring, then the $m=2$ mode will have completed two laps and so on. We call this the ``return time'' $t_R$ and derive it in \cref{app:GroupPhase}. Since the $\vb{A}$-field is constant, each lap results in the same Aharonov-Bohm phase, such that the time evolution from $t=0$ to $t=t_R$ results in
\begin{align}
    \psi_m(\phi) & \rightarrow \exp(m\frac{iq}{\hbar}\oint_0^{2\pi}A_\phi r\dd\phi)\psi_m(\phi)\:,\label{eq:ABmultipleTrips}
\end{align}
where the limits $0$ to $2\pi$ on the integral indicate that in this case the ring is only looped once. 
Due to the factor of $m$ in the exponent of \cref{eq:ABmultipleTrips} $\psi_m$ exactly picks up a Fourier multiplier of the form we are after. All we need to do is to identify 
\begin{align}
    \frac{q}{\hbar}\oint_0^{2\pi}A_\phi r\dd\phi =\Delta\phi= -\phi_u \:, \label{eq:phaseDifference}
\end{align}
and we have obtained exactly the desired Fourier multiplier $e^{-im\phi_u}$. 
Note that this happens, thanks to the geometric nature of the Aharonov-Bohm effect, simply because the ring is looped $m$ times!  An illustration of this is made in \cref{fig:ABFourierMult}b. (Alternatively, the Aharonov-Bohm phase \cref{eq:phaseDifference} can be obtained by explicit time evolution using the minimal coupling Hamiltonian, Eq. (\ref{HminCoupling}). The details of this calculation is given in \cite{SplitQPEAB}. Also note that the Aharonov-Bohm phase can take any value and that it is crucial to distinguish it from the flux quantization which can occur in superconducting rings.)

At the return time, $t=t_R$, each $\psi_m$ picks up the Fourier multiplier, $e^{-im\phi_u}$, and we get
\begin{align}
    \delta(\phi) \rightarrow \lim_{l\rightarrow\infty}\frac{1}{\sqrt{2\pi (2l+1)}}\sum_{m=-l}^le^{im(\phi-\phi_u)} = \delta(\phi-\phi_u)\:.
\end{align}
And voilá! We have moved information around a quantum system as desired in Eq.~(\ref{eq:deltaShift}) and can read out the phase $\phi_u$ by measuring the angular position of the particle.

Note that the fact that each of the particles Fourier basis states, $\psi_m$, picks up an $m$-dependent phase, $e^{-im\phi_u}$, is exactly the quantum parallelism that is often refered to when explaining how quantum computers work. Here the $m$-dependent phase is picked up simply because the state $\psi_m$ winds aounds the ring $m$ times in time $t_R$ (again the same result of course follows by doing the time evolution of the wave function explicitly, for details see \cite{SplitQPEAB}).

But wait! When we make the identification in  \cref{eq:phaseDifference}
don't we already know $\phi_u$? For now this is indeed the case.
As we explain below, because we have considered an electromagnetically charged particle on the ring with an electromagnetic field through the coil, this corresponds to $U$ being a $1\times1$ matrix. Obviously, the single matrix element in a $1\times1$ matrix is identical to its eigenvalue, so if we know $U$ we automatically also know $\phi_u$. In conclusion, the particle on the ring we have considered this far does work as a quantum computer which can solve our eigenvalue problem. But for now it can only do so for a $1\times1$ unitary matrix. Fortunately, as we shall now see, if we consider generalised vector potentials, our system with the particle on the ring also solves the eigenvalue problem, Eq. (\ref{eq:TheProblem}), for general $n\times n$ unitary matrices,  -- \emph{without} á priori knowledge of the eigenvalues. 

\subsection{Aharonov-Bohm Effect for $n\times n$ Matrices}\label{sec:nonAbelian}

The Aharonov-Bohm effect for an electromagnetic field depends on  the strength of the magnetic flux through the solenoid, ie.~a number. Consequently, the Aharonov-Bohm effect, in this case, is given by a complex phase, that is it represents the action of a $1\times1$ unitary matrix. However, to set up a quantum system (our quantum computer) which solves the full eigenvalue problem, \cref{eq:TheProblem}, all we need is to replace  $\vb{A}=A_\phi\hat{\phi}$ with an $
n\times n$ Hermitian matrix. The Aharonov-Bohm phase now becomes an $n\times 
n$ unitary operator \cite{PhaseFactors}
\begin{equation}
    \exp(\frac{iq}{\hbar}\int_\gamma \vb{A}\cdot \dd \vb{r})=\exp(\frac{iq}{\hbar}\oint A_\phi r\dd\phi)\:, \label{eq:nonAbelianPhase}
\end{equation}
since $\exp(iH)$ is unitary when $H$ is Hermitian.
 The electromagnetic potential is called a $U(1)$ \emph{gauge field}, since the phase it produces in \cref{eq:nonAbelianPhase} is part of the mathematical group $U(1)$, denoting the group of $1\times1$ unitary matrices, e.g. complex phases. In making the $\vb{A}$-field a Hermitian matrix it becomes a $U(n)$ gauge field since the ``phases'' generated in \cref{eq:nonAbelianPhase} are $n\times n$ unitary matrices.

 \subsection{The physical realisation of $U$ and $\ket{u}$}
 
In our setup with the particle on the ring (our quantum computer) the matrix $U$ in our original problem, \cref{eq:TheProblem}, is realised by
\begin{align}
    U = \exp(-\frac{iq}{\hbar}\oint_0^{2\pi} A_\phi r\dd\phi)\:,\label{eq:UnitaryPhaseRelation}
\end{align}
where again the boundaries on the integral indicates that the integral is for a single lap around the ring. The eigenvector $\ket{u}$ in \cref{eq:TheProblem} is realised as an internal degree of freedom for the particle on the ring. In the $n=1$ case, that internal degree of freedom is the electromagnetic charge of the particle, but for general $n$ we call it \emph{colour} \cite{YangMills,Color}. The  fundamental particles known as quarks have three colours, but mathematically there is nothing wrong with having $n$ colours \cite{Srednicki}. As such, we can represent $\ket{u}$ with the colour state of the particle. (We stress that this notion of quarks and colour is merely an analogy and we hope that for those familiar with the theory of the strong interactions this use of language can be useful.)

\subsection{The physical solution for $n\times n$ matrices}

Now it is natural to expect that the matrix nature of $U$ in Eq.~(\ref{eq:UnitaryPhaseRelation}) will complicate the realisation of the Fourier multipliers  tremendously, but because we consider the action of $U$ on an eigenstate $\ket{u}$ the matrix, $U$, can be replaced by its eigenvalue, $e^{i\phi_u}$, and everything is again just like in the electromagnetic case we considered above. To see this explicitly, let us describe the full state of a single mode as $\ket{m}\otimes\ket{u}$, where $\psi_m(\phi)=\braket{\phi}{m}$. If $\ket{u}$ is an eigenstate of $A_\phi$ with eigenvalue $-\hbar\phi_u/2\pi rq$ we get 
\begin{align}
    \ket{m}\ket{u} \rightarrow& e^{\frac{iq}{\hbar}\oint A_\phi r\dd\phi}\ket{m}\ket{u}\:, \label{eq:ABphaseEffect}\\
                    = & \ket{m}e^{\frac{iq}{\hbar}m2\pi r A_\phi}\ket{u}\:,\\
                    = & e^{-im\phi_u}\ket{m}\ket{u}\:, \label{eq:ABFourierMult}
\end{align}
and hence we obtain the desired Fourier multiplier (in the second line we use that $A_\phi$ is constant and independent of $\phi$). The Fourier multipliers shifts the particle exactly as in \cref{eq:deltaShift} and we can read out the phase $\phi_u$ by measuring the angular position of the particle. Let's stress that we do not need to know $\phi_u$ a priori, only that $\ket{u}$ is indeed an eigenstate of $U=\exp(-\frac{iq}{\hbar}\oint_0^{2\pi} A_\phi r\dd \phi)$, exactly the problem presented in \cref{sec:TheProblem}.  

In summary: 
When generalised to $U(n)$ gauge fields and $n$-colour particles, the quantum system made up of a particle on a ring with a coil through makes up a quantum computer capable of determining the phase of the eigenvalue 
 in \cref{eq:TheProblem}. The matrix $U$ is realised by the generalised Aharonov-Bohm phase, \cref{eq:UnitaryPhaseRelation}, and the eigenvector $\ket{u}$ is the colour state of the particle. Because the matrix $U$ acts on an eigenstate $\ket{u}$ it can be replaced by its eigenvalue corresponding to the eigenstate 
and the procedure outlined in \cref{sec:ABeffect} can be used to determine the phase $\phi_u$. The central quantum parallelism of the quantum computation happens simply because the $m$'th state, $\psi_m$,  loops the ring $m$ times in time $t_R$ and hence it picks up the Aharonov-Bohm phase factor $m$ times. 

For the reader interested non-Abelian gauge fields, be aware that we here have made some simplifications and choices of notation, as outlined in \cref{app:simple}. 

\section{An Explicit Example}\label{sec:ABExample}

The winding of the modes around the ring hopefully gives a physical intuition for the quantum parallelism which occurs in our quantum computation. However, the generalisation to $U(n)$ gauge fields adds a level of abstraction to this otherwise simple physical system. In order make this less abstract, we here provide an  example using a quark on the ring and the language of colour. The example, shows how this quantum system can be used to determine  an eigenvalue of a $2\times2$ unitary matrix. (As mentioned above quarks carry 3 colours so we could have considered a $3\times3$ matrix, but we limit ourselves to two colours for simplicity.) 
\bigskip

\noindent
{\sl The physical setup:} Suppose that instead of an electron, we place a quark on the ring, and instead of an electro-magnetic flux through the solenoid, there is now a colour-magnetic flux and a corresponding colour $\bf A$-field around the ring (borrowing from \cite{PreskillColour} one can imagine this flux being generated by a current of quarks in a coil around the solenoid, much like a current of electrons generates a magnetic field. For details on physical realisations of non-Abelian gauge fields, please, see \cite{DynamGauge,GaugeReal}).
\bigskip

\noindent
{\sl The problem:}
Let us retun to the example introduced in Section \ref{sec:TheProblem} and consider the unitary operator
\begin{align}
    U = e^{i\frac{H}{E_R}}\:,
\end{align}
where 
\begin{align}
H=E_0\begin{pmatrix}
        0 & 1 \\ 1 & 0
    \end{pmatrix} \ ,
\end{align}
and $E_0>0$. $E_R$ is a reference energy which sets the energy scale.
Just as for the general problem stated in Section \ref{sec:TheProblem}, we know $U$ and that 
\begin{align}
|u\rangle=\frac{1}{\sqrt{2}}\begin{pmatrix}
        1 \\ -1
    \end{pmatrix} \ ,
    \label{ex:eigenstate}
\end{align}
is an eigenstate of $U$. Our aim is to determine the phase, $\phi_u$, of the eigenvalue, $e^{i\phi_u}$, of $U$ associated with $\ket{u}$. (Note that in this example this corresponds to dertermining the ground state energy of $H$.) 
\bigskip

\noindent
{\sl How the problem is encoded in the physical system:} To encode this problem into our physical system, note that the quark has both an angular position wave function and a colour state $\ket{u}$. The colour state can be a superposition of red and blue, $\ket{u}=c_r\ket{r}+c_b\ket{b}$, where $c_r$ and $c_b$ are coefficients. We choose to work in the basis defined by $\ket{r}$ and $\ket{b}$, and the column vector \cref{ex:eigenstate} therefore can be encoded as
\begin{align}
   \ket{u}=\frac{1}{\sqrt{2}}(\ket{r}-\ket{b})\:.\label{eq:gaugeSimul}
\end{align}
The unitary matrix, in this example, is realised  by the $2\times2$ Aharonov-Bohm phase obtained as the quark loops the ring once
\begin{align}
    U= \exp({i\frac{H}{E_R}})&=\exp(-\frac{iq}{\hbar}\oint_0^{2\pi} A_\phi r\dd\phi)\:,
\end{align}
where
\begin{align}
    A_\phi=-\frac{\hbar}{q2\pi r}\frac{E_0}{E_R}\begin{pmatrix}
        0 & 1 \\ 1 & 0
    \end{pmatrix}\:.\label{eq:gaugeSimul2}
\end{align}
\bigskip
Note again that $\ket{u}$ is an eigenstate of $A_\phi$ and thus of $U$.

\noindent
{\sl How the problem is solved by the physical system:} We initialise our physical system at $t=0$ such that the quark is localised at $\phi=0$ and in the colour state $\ket{u}=\frac{1}{\sqrt{2}}(\ket{r}-\ket{b})$. Moreover, a constant quark current runs in the coil such that the A-field is given by \cref{eq:gaugeSimul2}.

As the time evolves from $0$ to $t_R$, the quark will remain in the colour state \cref{eq:gaugeSimul} and the $\vb A$-field will remain constant. However, the  position wave function of the quark, will spread out over the entire ring, as the individual Fourier modes with fixed angular momentum of the original $\delta$-function, start looping the ring. At the time $t_R$ the $m=1$ mode will have made one loop of the ring, and the modes with general $m$ will have made $m$ loops. Each loop adds an Aharonov-Bohm phase 
\begin{eqnarray}
     \exp(\frac{iq}{\hbar}\oint_0^{2\pi} A_\phi r\dd\phi)|m\rangle \ket{u} & = &  \exp(-im\frac{E_0}{E_R}\begin{pmatrix}
        0 & 1 \\ 1 & 0
    \end{pmatrix})|m\rangle \frac{1}{\sqrt{2}}\begin{pmatrix}
        1 \\ -1
    \end{pmatrix}\: ,
    \\ & = &
    \exp(-im\frac{E_0}{E_R}(-1))|m\rangle \frac{1}{\sqrt{2}}\begin{pmatrix}
        1 \\ -1
    \end{pmatrix}
     \:.
\end{eqnarray}
Note that in the first line the matrix which makes up the Aharonov-Bohm phase acts on one of its eigenstates, therefore we can replace the matrix by the associated eigenvalue, $e^{i\phi_u}=e^{-iE_0/E_R}$, in the second line. It is exactly at this step the information about the value of the phase $\phi_u$ (which we wish to determine) is extracted, without knowing $\phi_u$ a priori. Due to the factor of $m$ in the exponent, which simply comes from the fact that the $m$'th mode loops the ring $m$-times in  time $t_R$, the factor $e^{-im E_0/E_R}$ acts as a Fourier multiplier. The Fourier multipliers shift the angular position wave function of the quark, such that the quark at time $t_R$ is localised at $\phi=-E_0/E_R$. A measurement of the angular position of the quark at time $t_R$ therefore determines the phase of the eigenvalue $\phi_u=-E_0/E_R$ (and in this case the ground state energy of the Hamiltonian).

\begin{figure}
    \centering
    \includegraphics[width=0.9\linewidth,trim={0 0 0 12mm},clip]{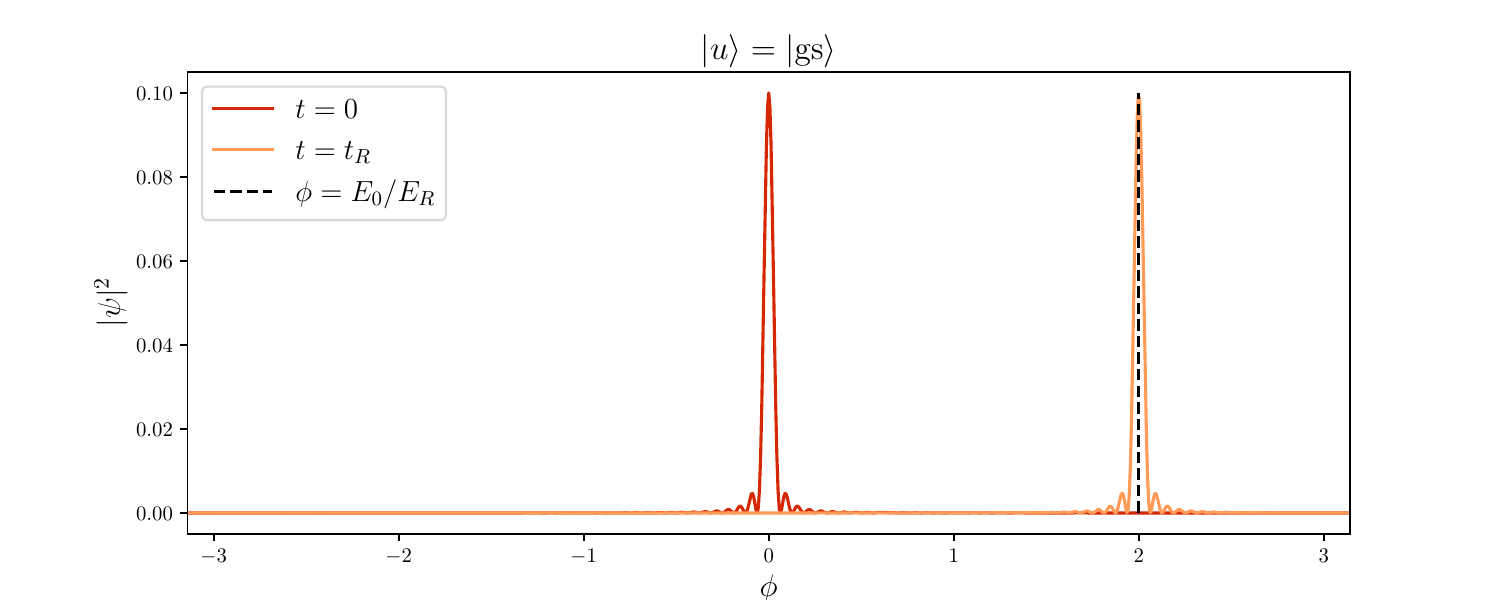}
    \caption{The probability density $\abs{\psi}^2$ for a particle on a ring experiencing a non-Abelian Aharonov-Bohm effect with $A_\phi$ and $\ket{u}$ given by \cref{eq:gaugeSimul2} and \cref{eq:gaugeSimul} for $E_0/E_R=2$. Simulated using the QuantumOptics.jl \cite{QuantumOpticsJL} library in Julia. The full Hamiltonian used to perform this simulation has size $200\times 200$, describing $100$ modes $\psi_m$, each with $2$ colours. Constructing the time evolution operator by exponentiation of the Hamiltonian requires $\mathcal{O}(200^2)$ operations at best and $\mathcal{O}(200^3)$ at worst on a classical computer \cite{MatrixExpScaling}, hence the classical method is inefficient.}
    \label{fig:ABsimulation}
\end{figure}

This example with $E_\textrm{0}/E_R=2$ is shown in \cref{fig:ABsimulation}. Of course, $2\times2$ matrices are easy to diagonalise by hand, this example acts more as a proof of concept that the quantum system can be used to determine eigenvalues. An example of what happens if $\ket{u}$ is \emph{not} an eigenstate of $A_\phi$ can be found in \cref{app:NonEig}.

\section{Quantum Phase Estimation}\label{sec:QPE}

So far we have shown a way of using our quantum system to determine eigenvalues of unitary operators. As for how this relates to quantum computing, this exact problem with this exact way of solving it is called \emph{Quantum Phase Estimation} \cite{kitaevQPE}. The QPE algorithm takes an $n\times n$ unitary operator $U$ and an eigenstate $\ket{u}$ and outputs an estimate of the phase of the eigenvalue $\phi_u$, solving the eigenvalue problem
\begin{equation}
    U\ket{u}=e^{i\phi_u}\ket{u}\:.
\end{equation}
QPE assumes access to some way of efficiently performing the operator $U$ on the state $\ket{u}$, and information about $\phi_u$ is extracted from this operation without any \emph{a priori} knowledge. QPE starts with initialising a quantum system of $t+\log_2n$ qubits, where $t$ is the number of decimal places to which we wish to estimate $\phi_u$. Typically we divide the qubits into \emph{registers}, with $t$ qubits in register 1 and $\log_2 n$ qubits in register 2. The procedure is then
\begin{equation}
    \overbrace{\ket{00\dots0}}^{t \textrm{ qubits}}\otimes \overbrace{\ket{u}}^{\log_2n \textrm{ qubits}} \xrightarrow{QPE \textrm{ and measurement}} \ket{k'}\otimes\ket{u}\:,
    \label{eq:QPEsimple}
\end{equation}
where, with high probability, $k'$ is a binary integer number such that $\abs{2\pi \frac{k'}{2^t} -\phi_u}<2^{-t}$, an approximation accurate to $t$ binary decimal places \cite{QuantAlgsRevisit}. Notice that only the qubits in register 1 change. Initially register 1 is all $0$'s but quantum phase estimation ``writes'' an approximation of $\phi_u$ into the register accurate to $t$ (binary) decimal places. Note that inherent within the ``measurement'' is a probability of failure, i.e. the integer in register 1 does \emph{not} correspond to the best $t$-decimal approximation of $\phi_u$. The success probability can be described as \cite{NielsenChuang}
\begin{align}
    p(\abs{k'-\mu}>e)=\frac{1}{2(e-1)}\:,
\end{align}
where $k'$ is the measured integer in register 1, $2\pi\frac{\mu}{2^t}$ is the best possible $t$ decimal approximation to $\phi_u$ and $e$ is some desired tolerance.
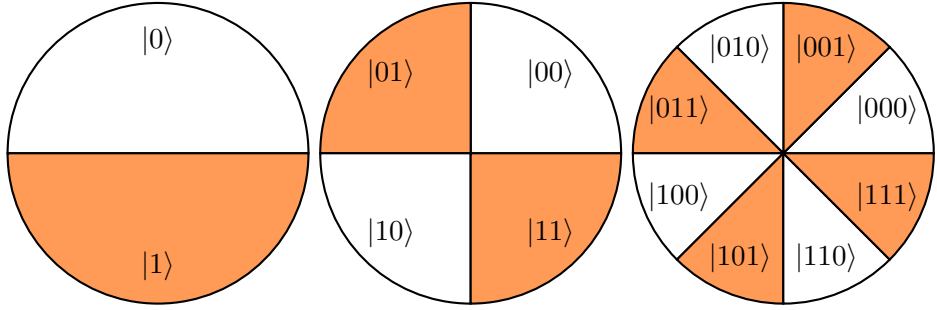
\begin{figure}
    \centering
    \begin{tikzpicture}
        \def\tikzset#1{\pgfkeys{/tikz/.cd,#1}};
        \def\numslices{2};
        \def\offset{360/\numslices};
        \def\size{2pt};
        \def\numdigits{1};
        \foreach \i in {1,...,\numslices} {
            \pgfmathtruncatemacro\myresult{mod(\i,2)};
            \ifnum\myresult>0
                \def\col{white};
            \else
                \def\col{lesbian-orange};
            \fi
            \def\angle{\i*\offset};
            \def\prevAngle{\angle-\offset};
            \draw[thick,fill=\col] (0cm,0cm)--(\angle:2cm) arc(\angle:\prevAngle:2cm)--cycle;
            \draw (\angle-\offset/2:1.5cm) node {$\ket{\pgfmathsetbasenumberlength{\numdigits} \pgfmathparse{bin(\i-1)} \pgfmathresult}$};
        }
    \end{tikzpicture}
    \begin{tikzpicture}
        \def\tikzset#1{\pgfkeys{/tikz/.cd,#1}};
        \def\numslices{4};
        \def\offset{360/\numslices};
        \def\size{2pt}
        \def\numdigits{2}
        \foreach \i in {1,...,\numslices} {
            \pgfmathtruncatemacro\myresult{mod(\i,2)}
            \ifnum\myresult>0
                \def\col{white}
            \else
                \def\col{lesbian-orange}
            \fi
            \def\angle{\i*\offset};
            \def\prevAngle{\angle-\offset}
            \draw[thick,fill=\col] (0cm,0cm)--(\angle:2cm) arc(\angle:\prevAngle:2cm)--cycle;
            \draw (\angle-\offset/2:1.5cm) node {$\ket{\pgfmathsetbasenumberlength{\numdigits} \pgfmathparse{bin(\i-1)} \pgfmathresult}$};
        }
    \end{tikzpicture}
    \begin{tikzpicture}
        \def\tikzset#1{\pgfkeys{/tikz/.cd,#1}};
        \def\numslices{8};
        \def\offset{360/\numslices};
        \def\size{2pt}
        \def\numdigits{3}
        \foreach \i in {1,...,\numslices} {
            \pgfmathtruncatemacro\myresult{mod(\i,2)}
            \ifnum\myresult>0
                \def\col{white}
            \else
                \def\col{lesbian-orange}
            \fi
            \def\angle{\i*\offset};
            \def\prevAngle{\angle-\offset}
            \draw[thick,fill=\col] (0cm,0cm)--(\angle:2cm) arc(\angle:\prevAngle:2cm)--cycle;
            \draw (\angle-\offset/2:1.5cm) node {$\ket{\pgfmathsetbasenumberlength{\numdigits} \pgfmathparse{bin(\i-1)} \pgfmathresult}$};
        }
    \end{tikzpicture}

    \caption{Intervals of positions along the ring for one, two and three qubits. The coloured slices are those with $1$ as the rightmost digit.}
    \label{fig:circPositions}
\end{figure}

We can think of the possible states of register 1 as a ring cut into $2^t$ slices. For example, with three qubits (eight slices) the state $\ket{000}$ would be the slice between $0$ and $\pi/4$, while $\ket{100}$ is the slice between $\pi$ and $5\pi/4$. The more qubits in register 1, the more slices the ring can be divided into, the better the precision of the output. This concept is illustrated in \cref{fig:circPositions}. With this visual in mind, we turn to describing QPE. 

The QPE algorithm \cite{kitaevQPE,QuantAlgsRevisit} consists of the four steps (after initialization) summarized in the table below \cite{NielsenChuang,SplitQPEAB}:
\begin{center}
\begin{tabular}{ l l}
{\bf Operation} & {\bf State}  \\
\hline
 Initial state & $|0\rangle|u\rangle$  \\ 
 Quantum Fourier transform & $\frac{1}{\sqrt{2^t}}\sum_{m=0}^{2^t-1}|m\rangle|u\rangle$  \\  
 Controlled applications of U &  $\frac{1}{\sqrt{2^t}}\sum_{m=0}^{2^t-1}e^{i\phi_u m}|m\rangle|u\rangle$  \\
 Inverse Quantum Fourier transform & $\frac{1}{2^t}\sum_{m,k=0}^{2^t-1}e^{i(\phi_u-2\pi\frac{k}{2^t}) m}|k\rangle|u\rangle$\\
 Measurement & $|k'\rangle|u\rangle$  \\
 \hline
\end{tabular}
\end{center}
Note that the state after the controlled applications of $U$ in the third line, has exactly the Fourier multipliers we discussed above. The final measurement therefore gives an accurate estimate $2\pi k'/2^t$ for the phase. A more detailed mathematical description of the algorithm can be found in literature \cite{NielsenChuang,qubitguide,QuantAlgsRevisit}.

\begin{figure}
    \centering
    \begin{tikzpicture}
        \graph [grow right sep] {"$\psi(\phi)$"[terminal] -> FT[nonterminal] -> "$c_m\psi_m$"[terminal] -> Fourier Multiplier[nonterminal] -> "$e^{-im\phi_u}c_m\psi_m$"[terminal] -> IFT[nonterminal] -> "$\psi(\phi-\phi_u)$"[terminal]};
    \end{tikzpicture}\vspace{5mm}
    \begin{tikzpicture}
        \graph [grow right sep] {"$\ket{0}\ket{u}$"[terminal] -> QFT[nonterminal] -> "$\ket{m}\ket{u}$"[terminal] -> Controlled Unitary[nonterminal] -> "$e^{im\phi_u}\ket{m}\ket{u}$"[terminal] -> IQFT[nonterminal] -> "$\ket{k'}\ket{u}$"[terminal]};
    \end{tikzpicture}
    \caption{{\bf Top:} Diagram of using Fourier multipliers to shift the argument of a function. FT and IFT refer to the Fourier transform and its inverse. {\bf Bottom:} Simplified diagram of Quantum Phase Estimation. QFT and IQFT refer to the Quantum Fourier Transform and its inverse.}
    \label{fig:SimpleDiagram}
\end{figure}
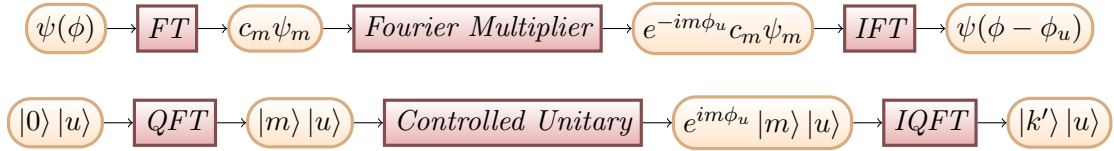

We can now compare this standard form of the QPE algorithm to that obtained using the Aharonov-Bohm phase.
\begin{enumerate}
    \item  \emph{Quantum Fourier transform} \\ \textbf{QPE algorithm:} Create an even superposition of all possible states in register 1. 
    \begin{align}
        \ket{00\dots0}\rightarrow\frac{1}{\sqrt{2^t}}\sum_{m=0}^{2^t-1}\ket{m}\:.
    \end{align}
    \textbf{Aharonov-Bohm approach:} this transformation corresponds to expanding the initial wave function on the stationary states of the Hamiltonian, so this operation is ``free'' in the physical system. 
    \item \emph{Controlled applications of $U$} \\ \textbf{QPE algorithm:} Apply the operator $U^{2^{j}}$ to register 2 if the $j$'th qubit in register 1 is in the state $\ket{1}$. Notice that ``if'' on a quantum computer includes superpositions, entangling the states of registers 1 and 2. 
    We can write these operations explicitly as $cU = \sum_m\ketbra{m}{m}\otimes U^{m}$. \\
    \textbf{Aharonov-Bohm approach:} This is exactly analogous to the effect of the Aharonov-Bohm phase in \cref{eq:ABphaseEffect}.
    \item \emph{Inverse Quantum Fourier transform} \\ \textbf{QPE algorithm:} Invert the Quantum Fourier transform. If register 1 had an infinite amount of qubits this would give us a state perfectly localised at $\phi_u$, just like in \cref{eq:inverseFourier}. However, for finite number of qubits we get a state which is narrowly peaked near $\phi_u$, as in \cref{eq:QPEsimple} \cite{QuantAlgsRevisit,NielsenChuang}. \\
    \textbf{Aharonov-Bohm approach:} Once again, we get this transformation ``for free'' in the particle on a ring system (see the final step below).
    \item \emph{Measurement} \\ \textbf{QPE algorithm:} Measure the state of each qubit in register 1. This measurement will result in a string of $1$'s and $0$'s: a binary integer $k'$ of length $t$. The output will (with high probability) be such that $\abs{\phi_u-2\pi \frac{k'}{2^t}}\leq2^{-t}$, e.g. $2\pi \frac{k'}{2^t}$ will be an approximation of $\phi_u$ correct to $t$ decimal places in binary. \\
    \textbf{Aharonov-Bohm approach:} For the particle on the ring, measuring the position automatically projects on to the Fourier inverse of the definite angular momentum states $\psi_m$. The measured position gives the desired value of $\phi_u$.
\end{enumerate}
As detailed the procedure of quantum Fourier transforms and Fourier multipliers is exactly the same in the QPE algorithm as in the Aharonov-Bohm picture we described earlier. The system and the algorithm are therefore equivalent, save for the implicit discreteness of the possible positions on the algorithm side, see \cref{fig:SimpleDiagram}.

QPE 
enters as a subroutine in Shor's algorithm for prime factorisation \cite{Shor} and the Harrow-Hassidim-Lloyd algorithm for solving linear systems of equations \cite{HHL}, both of which can provide exponential speedup over the best know classical algorithms \cite{Shor,HHL}.

\section{Simulation}\label{sec:simul}

A particle on a ring is seemingly a very simple quantum systems. Could QPE be performed efficiently on classical hardware by numerical simulation of this system? The short answer is \emph{no}, but exploring why illustrate the power of quantum computers. We will perform the simulation with a brute force approach, constructing the time evolution operator explicitly. The main problem here is the numerical precision. Say that we wish to estimate $\phi_u$ to $t$ (binary) decimal places with a success probability of $1-\epsilon$, i.e.
\begin{align}
    P(k')\geq1-\epsilon,\quad\abs{\phi_u-\frac{2\pi k'}{2^t}}\leq2^{-t}\:.
\end{align}
To even resolve this precision we need $N\ge 2^t$ points along the ring, meaning that the necessary size of the Hilbert space needed to simulate the QPE algorithm on a classical computer using this method grows as $\mathcal{O}(2^{ t})$ with the binary precision $t$. The simulation is performed by application of the operator $e^{-iHt_R/\hbar}$ with 
\begin{align}
    H = \frac{1}{2m_q}\left(-\frac{i\hbar}{r} \pdv{}{\phi}-qA_\phi\right)^2\:.
    \label{HminCoupling}
\end{align}
Since this operator acts on both momentum and colour space, it is of size $(N\cdot n)\times (N\cdot n)$. Constructing the exponential of an operator using the packages provided in the Julia programming language requires $\mathcal{O}((N\cdot n)^2)$ operations at best and $\mathcal{O}((N\cdot n)^3)$ at worst \cite{MatrixExpScaling}, i.e. $\mathcal{O}(2^{3t}n^3)$. Compare this to QPE on a quantum computer, which requires $t$ applications of operators of the type $U^{2^j}$, which are all $n\times n$, as well as $\mathcal{O}(t^2)$ operations for QFT \cite{QuantAlgsRevisit}. Thus, simulating the Aharonov-Bohm effect for a particle on a ring using a classical computer is highly inefficient compared to QPE, as illustrated by the example shown in \cref{fig:ABsimulation}.

\section{Conclusion} 

It can be difficult for students to enter the field of quantum computation because quantum computers operate in a manner so different from classical computers that our intuitions for understanding the process of quantum computation is challenged.
The aim of this article is to help students build a physical intuition for quantum algorithms. 
A quantum computer is a quantum system and starting from the perhaps simplest quantum system, the particle on a ring, we showed how the Aharonov-Bohm effect can be used to determine eigenvalues of unitary matrices, in a way which is equivalent to the central Quantum Phase Estimation algorithm. Said in other words the particle on the ring system make up a quantum computer capable of performing quantum phase estimation. For this specific quantum computer, the often mysterious quantum parallelism of the quantum computation, is realised simply because the modes with fixed angular momentum, which makes up the position wave function of the particle, loops the ring a number of times which equals the angular momentum quantum number. For each loop the mode picks up an Aharonov-Bohm phase factor, and these add up to a Fourier multiplier which shifts the angular position wave function. The shift equals the phase of the eigenvalue which is to be computed and the result can be read out by measuring the angular position of the particle.    

Lastly, we discussed simulating the particle on the ring (our quantum computer) numerically on a classical computer. While the system may appear simple, we showed that the amount of operations needed to simulate it scales poorly compared to the Quantum Phase Estimation quantum algorithm. This illustrates the power of quantum computing. Indeed if it was possible to simulate this system effectively by classical methods there would be no advantage to use a quantum computer for the estimation of the phase of the eigenvale of the unitary operator.  

We have presented the example of using the Aharonov-Bohm effect for a particle on a ring to determine eigenvalues of unitary matrices in lectures at a graduate level quantum mechanics course and at a PhD school. Both groups of students responded positively to the example, claiming that it helped develop an intuition for quantum algorithms. Our hope is that this article can perhaps also serve as inspiration for other lecturers of graduate level courses on quantum mechanics and offer them a possible way to integrate an introduction to quantum algorithms into their courses. 

\section{Acknowledgments}
We thank Bjarke T. R. Christensen for discussions and Victoria Inselmann for comments. This work is supported by the Novo Nordisk Foundation, Grant number NNF22SA0081175, NNF Quantum Computing Programme.

\section{Conflicts of Interest}
The authors have no conflicts to disclose.

\section{Author Contributions}
A. L. Bjerregaard wrote the initial manuscript, supervised by K. Splittorff. Both authors contributed equally to the editorial process.

\appendix
\crefalias{section}{appendix}

\section{Phase vs Group Velocity}\label{app:GroupPhase}

Any possible wave function $\Psi(t,\phi)$ for a particle on a ring evolving under the Hamiltonian
\begin{align}
    H = \frac{L^2}{2I}\:,\quad I=m_qr^2
\end{align}
is time periodic, $\Psi(t+t_R,\phi)=\Psi(t,\phi)$. The \emph{return time} $t_R$ can be found by writing out $\Psi(t,\phi)$ explicitly \cite{SplitQPEAB}. (Note that we here for simplicity consider the case of a vanishing magnetic flux. The derivation with a non-zero magnetic flux follow the exact same lines, see \cite{SplitQPEAB}.)
\begin{align}
    \Psi(t,\phi)&=\lim_{l\rightarrow\infty} \frac{1}{\sqrt{2l+1}}\sum_{m=-l}^lc_m\psi_m(\phi)e^{-\frac{iE_mt}{\hbar}}\:,\\
    \psi_m(\phi)&=\frac{1}{\sqrt{2\pi r}}e^{im\phi},\quad E_m = \frac{\hbar^2m^2}{2m_qr^2}\:,\\
     \Psi(t+t_R,\phi)&=\lim_{l\rightarrow\infty} \frac{1}{\sqrt{2l+1}}\sum_{m=-l}^lc_m\psi_m(\phi)e^{-\frac{iE_mt}{\hbar}}e^{-\frac{iE_mt_R}{\hbar}}\:.
\end{align}
For $t_R=4\pi m_qr^2/\hbar$ the factors $e^{-\frac{iE_mt_R}{\hbar}}=e^{-i2\pi m^2}$ all evaluate to $1$ and thus $\Psi(t+t_R,\phi)=\Psi(t,\phi)$. The individual modes $\psi_m$ are eigenstates of the angular momentum operator $L$ with $L\psi_m=\hbar m \psi_m$. From this we could define a sort of ``classical'' angular velocity for each mode as
\begin{align}
    \omega_m^{\textrm{cl}}=\frac{\hbar m}{I}=\frac{\hbar m}{m_qr^2}\:.
\end{align}
However, comparing with the return time we just calculated we find $\omega_m^{\textrm{cl}}t_R=4\pi m$, suggesting that each particle has some underlying ``quantum'' angular speed with the relation
\begin{align}
    \omega^{\textrm{cl}}=2\omega^{\textrm{qm}}\:.
\end{align}
Notice that this is completely analogous to phase and group velocity in free particle wave packets on $(-\infty,\infty)$, where $v_\textrm{group}=2v_\textrm{phase}$ \cite{Griffiths}. 

\section{Non-Eigenvalue Colour Configuration}\label{app:NonEig}

\begin{figure}
    \centering
    \includegraphics[width=0.9\linewidth,trim={0 0 0 12mm},clip]{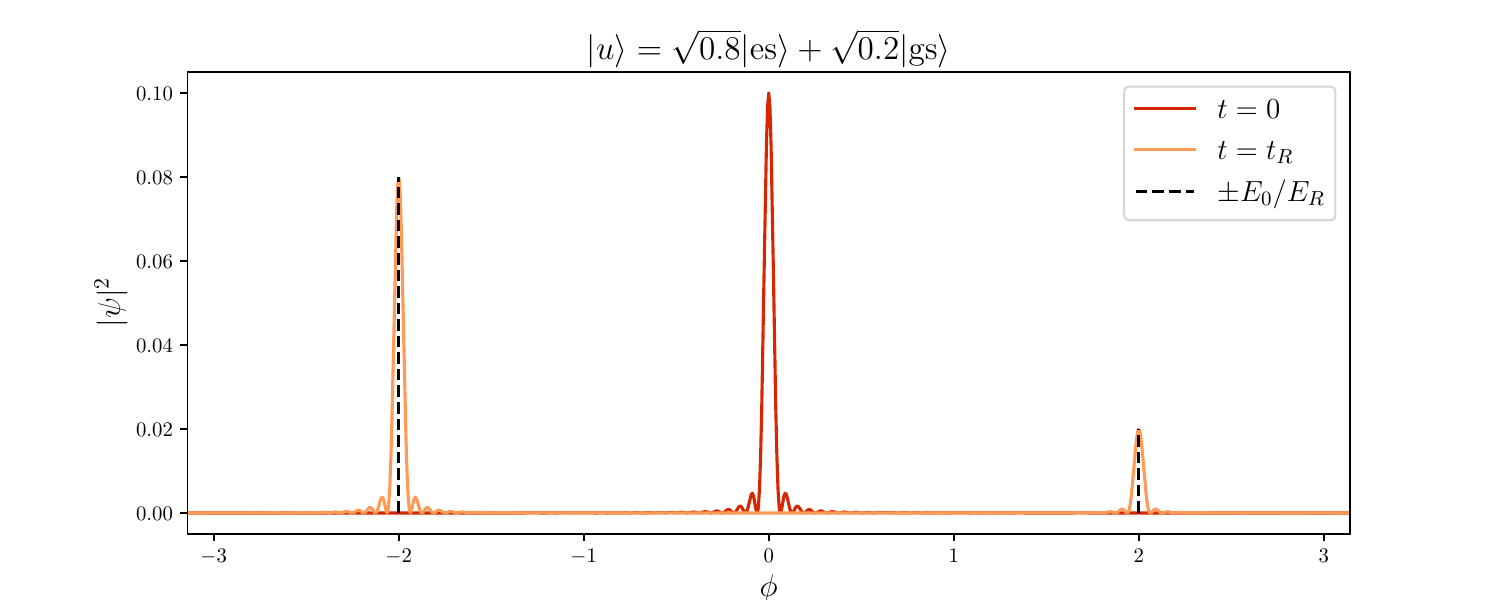}
    \caption{An initially localised wave function which is in a superposition of two eigenstates of $U$ will develop two peaks at time $t_R$. For the example we have chosen $E_0/E_R=2$ with $A_\phi$ and $\ket{u}$ given by \cref{eq:multipleEigVecs}.}
    \label{fig:ABnonEig}
\end{figure}

What happens to our particle on a ring if the colour state $\ket{u}$ is \emph{not} an eigenstate of the gauge field $A_\phi$? Suppose $A_\phi$ has eigenstates $\ket{k},k=1,\dots,n$. Since the gauge field is Hermitian, we can expand $\ket{u}$ in the basis of eigenstates,
\begin{align}
    \ket{u}=\sum_{k=1}^nc_k\ket{k}\:.
\end{align}
In this basis, we can describe how the Aharonov-Bohm phase will act on the colour space. The Aharonov-Bohm phase acts linearly, such that for a single lap around the ring the colour space is rotated by
\begin{align}
    \exp(i2\pi r\frac{q}{\hbar}A_\phi)\ket{u}=\sum_{k=1}^nc_k\exp(i2\pi r\frac{q}{\hbar}\lambda_k)\ket{k}\:,
\end{align}
where $\lambda_k$ satisfies $A_\phi\ket{k}=\lambda_k\ket{k}$. Letting $\phi_k = 4\pi r\frac{q}{\hbar}\lambda_k$, with the extra factor of $2$ from the discrepancy between $\omega^\textrm{cl}$ and $\omega^\textrm{qm}$, an initially localised particle will transform as
\begin{align}
    \delta(\phi)\otimes\ket{u}&\xrightarrow{t=t_R}\lim_{l\rightarrow\infty}\frac{1}{\sqrt{2l+1}}\sum_{m,k}c_ke^{im(\phi-\phi_k)}\otimes\ket{k}\:,\\
    &=\sum_{k=1}^nc_k\delta(\phi-\phi_k)\otimes \ket{k}\:.
\end{align}
So for $\ket{u}$ being a superposition of eigenstates $\ket{k}$, the particle will re-localise at each phase $\phi_k$ with an amplitude proportional to $\abs{c_k}^2$. Returning to our example in \cref{sec:ABExample}, if we let
\begin{align}
    A_\phi=-\frac{\hbar}{q2\pi r}\frac{E_0}{E_R}\begin{pmatrix}
        0 & 1 \\ 1 & 0
    \end{pmatrix}\:, \quad {\rm and} \quad
    \ket{u}=\sqrt{0.8}\frac{1}{\sqrt{2}}\begin{pmatrix}
         1 \\ -1
    \end{pmatrix}+\sqrt{0.2}\frac{1}{\sqrt{2}}\begin{pmatrix}
         1 \\ 1
    \end{pmatrix},\label{eq:multipleEigVecs}
\end{align}
we will get a wave function with two peaks at $t=t_R$. The amplitudes of the peaks correspond to the absolute square of the projection of $\ket{u}$ onto the eigenstates of $U$, see \cref{fig:ABnonEig}. 

\section{Simplifications}\label{app:simple}

Compared to how non-Abelian gauge fields are normally introduced, we have made some simplifications in this article. Generally, the Aharonov-Bohm phase for a non-Abelian gauge field should be written as \cite{PhaseFactors}
\begin{align}
    \mathcal{P}\exp(\frac{iq}{\hbar}\int_\gamma \vb{A}\cdot \dd\vb{r})\:,
\end{align}
where $\mathcal{P}$ denotes \emph{path ordering}. However, as $\vb{A}=A_\phi\hat{\phi}$ is constant in our physical setup, path ordering is unnecessary.

Concerning the notation we have used for the non-Abelian gauge field: Normally one makes the special relativistic nature of the theory apparent by the use of four-vectors. The same can be done for the vector potential $\vb{A}(\vb{r})\rightarrow A_\mu,\mu=0,1,2,3$. In the electromagnetic case $A_0$ represents the electric potential while $A_{1,2,3}$ are the components of the magnetic vector potential. In the case of non-Abelian gauge fields we typically write them as $A_\mu = A_\mu^aT_a,a=0,\dots,N^2-1$, where $T_a$ is a basis for all $n\times n$ Hermitian matrices. Here, the repeating of the index $a$ indicates summation over it. Thus for every value of $\mu$, $A_\mu$ is a sum of Hermitian matrices with coefficients $A_\mu^a$. Often one works directly with these coefficients, calling $\mu$ the \emph{spacetime} index, since it denotes which direction in spacetime we are considering, and $a$ the \emph{colour} index, denoting the internal colour space direction. However, since our $\bf{A}(\bf{r})$ is restricted to the $\hat{\phi}$ direction, we chose to keep the  vector notation to more easily express the non-Abelian phases as Aharonov-Bohm phases.

\end{document}